\documentclass[aps,prb,twocolumn,groupedaddress,floatfix]{revtex4-1}
\usepackage{amsmath}
\usepackage{amsfonts}
\usepackage{amssymb}
\usepackage{graphicx}
\usepackage{bm}
\usepackage{bbold}
\usepackage{float}
\usepackage{hyperref}

\begin{document}
\title{Majorana and fractionally charged bound states in 1-D  Rashba nanowire under spatially varying Zeeman fields}
\author{Xiaoyu Zhu}
\affiliation{National Laboratory of Solid State Microstructures and Department of Physics, Nanjing University, Nanjing 210093, China}
\author{Wei Chen}
\affiliation{National Laboratory of Solid State Microstructures and Department of Physics, Nanjing University, Nanjing 210093, China}
\author{R. Shen}
\affiliation{National Laboratory of Solid State Microstructures and Department of Physics, Nanjing University, Nanjing 210093, China}
\author{D. Y. Xing}
\affiliation{National Laboratory of Solid State Microstructures and Department of Physics, Nanjing University, Nanjing 210093, China}

\date{\today}

\begin{abstract}
We study topological phase transitions in one dimensional (1-D) Rashba nanowire under a spatially varying Zeeman field when coupled to an $s$-wave superconductor substrate. We show that this system supports both Majorana bound states (MBS) and fractionally charged bound states (FBS) of Jackiw-Rebbi type. By disassembling Zeeman Hamiltonian into multiple helical components, we find that each helical component is relating to a corresponding topological region, characterized by the emergence of MBS. FBS arises in the overlapping gapped area created by any two helical components with topologically differing configuration, analogous to those formed at the knot in SSH model. We then develop a general criteria for the occurrence conditions of MBS and FBS. Our results suggest that systems with large Rashba spin orbit couple amplitude or in presence of weak Zeeman fields favor MBS, and otherwise FBS are more favorable. In the end, we demonstrate that spin components of zero energy bound states in topological phases are polarized in the plane perpendicular to Rashba vector, and the polarization oscillates with the variance of phases of the Zeeman field.
\end{abstract}

\pacs{}

\maketitle
\section{introduction}\label{sec:introduction}
Majorana fermions, which are their own anti-particles,\cite{Majorana} have recently been predicted to exist in condensed matter systems as Majorana bound states (MBS) at zero energy levels.\cite{Wilczek-MF,franz-race,franz_majorana_wire,Beenakkersearch,AliceaReview} The pursuit of MBS has twofold significance, both for the fundamental physics and for the implementation of fault-tolerant topological quantum computation due to their non-Abelian statistics.\cite{Kitaev,Read/paired,Moore/nonabelion,Non-Abelian-RMP,Alicea/natpy} Until now, a number of proposals supporting MBS have been put forward,\cite{Fu/proximity,Lee-MF-FS,Sau/generic,Alicea/tunable,Cook-MF-QH,Lutchyn/1d-semiconductor,Oreg/helical-liquids} one of which is utilizing Rashba nanowire coupled to an $s-$wave superconductor in presence of a uniform Zeeman field.\cite{Cook-MF-QH,Alicea/tunable,Lutchyn/1d-semiconductor,Oreg/helical-liquids} With the observation of MBS signals in hybrid nanowire-superconductor devices,\cite{Mourik,das/zero,dengmt/anomalous, Rokhinson} the latter proposal seems to be much closer to experimental verification of MBS.

It was suggested that a helical Zeeman field is equivalent to the combination of Rashba spin orbit coupling (RSOC) and a uniform Zeeman field in a one dimensional (1-D) nanowire,\cite{Braunecker/spin-selective} indicating that MBS may emerge without RSOC when a helical field is applied.\cite{Beenakker/without-spin-orbit,Flensberg/without-spin-orbit} Further steps have been made when systems favoring helically arranged localized magnetic moments were proposed,\cite{Klinovaja/RKKY,Franz/self-organized,Braunecker/interplay} which are able to self-tune themselves into topological phases, since the spatial period of the magnetic moment array is determined by Fermi wave vector. Moreover, it was demonstrated that, not only MBS, but fractionally charged bound states (FBS), can possibly appear when an extra uniform Zeeman field is applied in addition to the helical one.\cite{Localized_Gangadharaiah,fractionalfermions,Fractional-non-abelian} FBS are localized states resembling that in Jackiw-Rebbi model,\cite{Jackiw-Rebbi} and carry fractional charge. Under a helical field, energy levels of them behave sensitively with phases of the field.

Previous works mentioned above actually all focus on helical fields, $i.e.$, fields with only one unique helical component. Most spatially varying fields, however, consist of more than one helical components and the phase transition behaviors in presence of them is still not clear. In this paper, we shall study this general case in a unified framework. In order to understand the roles of the spatially varying fields, we first analyze the energy spectrum in absence of superconductivity. Our finding suggests that interplay between RSOC and Zeeman fields opens up multiple gaps at the edge of first Brillouin zone (FBZ). When chemical potential is tuned into any of these gaps, the system is expected to possess nontrivial topology, with MBS emerging at each end of the nanowire when $s$-wave pairing potential is turn on. Therefore, the phase diagram is expected to exhibit multiple topological windows. Employing numerical analysis, we also find that, with the increase of Zeeman fields, the system can be driven into phases supporting FBS, due to the strong interplay among different helical components of the fields. We then provide a general criterion to determine when MBS and FBS are supposed to appear. In the end, we calculate spin polarization profile of zero mode bound states in topological phases and our results indicate that spin is polarized only in the plane perpendicular to RSOC vector, and that spin polarization changes remarkably with phases of the Zeeman field.

This paper is organized as follows. In Sec.\ref{sec:continuum model}, we introduce the continuum model of our system and address topological phase boundaries under the spatially varying Zeeman field. In Sec.\ref{sec:numerical results}, we resort to numerical analysis to investigate MBS and FBS by calculating local density of states (LDOS), after which phase diagram is presented as well as spin polarization of zero energy bound states. Finally, summary is presented in Sec.\ref{sec:summary}.

\section{continuum model}{\label{sec:continuum model}}
\begin{figure}
\includegraphics[scale=0.37]{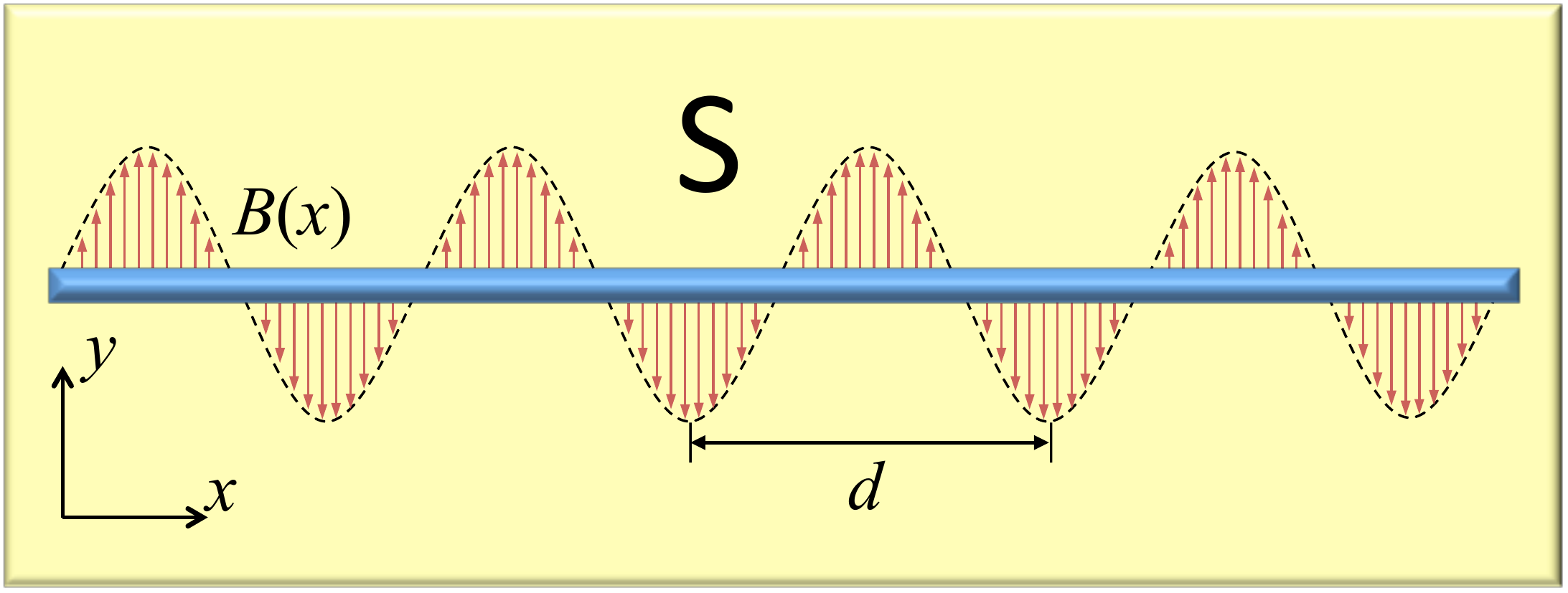}
\caption{(Color online) Sketch of the proposed setup. A nanowire is deposited on an $s$-wave superconductor. The spatially varying Zeeman field, which is aligned in $x-y$ plane, drives the nanowire into topological phases.}\label{fig:schematic}
\end{figure}

Our setup is sketched in Fig. \ref{fig:schematic}, where a 1-D nanowire with RSOC is in proximity with
an $s$-wave superconductor and subject to a spatially varying Zeeman field. Assume the wire lies in x direction and RSOC vector points along z direction. Since the Zeeman field parallel to RSOC vector cannot open a gap, we only consider fields applied in $x-y$ plane. BdG Hamiltonian can then be written in Nambu spinor basis $\{\psi_\uparrow(x),\psi_\downarrow(x),\psi_\downarrow^\dagger(x),-\psi_\uparrow^\dagger(x)\}$, as follows
\begin{equation}
H=(-\frac{\hbar^2\partial_x^2}{2m}-\mu)\tau_z-iA\partial_x\sigma_z\tau_z+\bm{V}(x)\cdot\bm{\sigma}+\Delta\tau_x, \label{eq:Hamiltonian}
\end{equation}
where $\bm{\sigma}=\{\sigma_x,\sigma_y,\sigma_z\}$ acts in spin space, and $\bm{\tau}=\{\tau_x,\tau_y,\tau_z\}$ in particle-hole space. In our model, $m$ is the effective band mass of an electron, $\mu$ the chemical potential, $A$ the strength of RSOC, and $\bm{V}(x)$ is the spatially varying Zeeman field. Experimentally, the spatially varying Zeeman fields can be generated by intrinsic nuclear spin, \cite{Braunecker_nuclear} or by external nanomagnet arrays.\cite{Karmakar} The Zeeman field with period $d$ can be written as the sum of Fourier series,
\begin{equation}
\bm {V}(x)=\sum\limits_{n;i=x,y} V_{ni}\cos(nK_dx+\phi_{ni})\bm{\hat{e}}_i
\end{equation}
where $K_d=2\pi/d$, is the reciprocal wave vector of the field, and $\phi_{ni}$ represents phases of the field in $i$ ($x$ or $y$) direction. Zeeman Hamiltonian is given by
\begin{equation}
H_Z(x)=\sum\limits_{n;\sigma=+,-}V_{n\sigma}e^{i\sigma nK_dx}\psi_\uparrow^\dagger(x)\psi_\downarrow(x)+\text{H.c.}\label{eq:Zeeman}
\end{equation}
where $V_{n\sigma}$ depends on $V_{ni}$ and $\phi_{ni}$. We call $V_{n\sigma}e^{i\sigma nK_dx}$ as the helical component of $\bm{V}$, since one such term is in fact equivalent to a helical Zeeman field. 

At first, we consider Zeeman fields with only two helical components, given by
\begin{equation}
\bm{V}(x)=V_x \cos(K_dx+\phi_x)\bm{\hat{e}}_x+V_y \sin(K_dx+\phi_y)\bm{\hat{e}}_y\label{eq:field}
\end{equation}
We not that results obtained in this case can be easily generalized to cases where Zeeman fields have more than two helical components. The corresponding Zeeman Hamiltonian of Eq.(\ref{eq:field}) is written as
\begin{equation}
H_Z(x)=(V_+e^{iK_dx}+V_-e^{-iK_dx})\psi_\uparrow^\dagger(x)\psi_\downarrow(x)+\text{H.c.}\label{eq:general-Zeeman}
\end{equation}
where $V_\pm=\frac{1}{2} e^{\pm i\phi_x}(V_x \mp V_ye^{\pm i\Delta\phi})$, representing the amplitude of the two helical components separately, and $\Delta\phi=\phi_y-\phi_x$, being the phase difference in x and y direction. It should be pointed that, FBZ is relocated in $[-K_d/2,K_d/2]$ under the spatially varying field. To obtain the energy spectrum, we first transform Eq.(\ref{eq:general-Zeeman}) into $k-$space,
\begin{equation}
H_Z(k)=\sum\limits_k (V_+c_{k\uparrow}^\dagger c_{k+K_d\downarrow}+V_-c_{k\downarrow}^\dagger c_{k+K_d\uparrow})+\text{H.c.}\label{eq:k-Zeeman}
\end{equation}
As usual, we fold the energy spectrum into FBZ and only consider energy bands in this zone. The operator $c_{k+n'K_d\uparrow(\downarrow)}$ can thus be rewritten as $c_{n'k\uparrow(\downarrow)}$, with $n'$ being the band index. Without Zeeman and pairing potential term, energy spectrum is given by
\begin{equation}
E_{n'\sigma}(k)=\frac{\hbar^2(k+n'K_d)^2}{2m}-\sigma A(k+n'K_d) \label{eq:band-dispersion}
\end{equation}
with $\sigma=\pm$, representing spin-up and spin-down respectively. Note that there are actually numerous degenerate energy levels at Kramer's degenerate points, $k=0$ and $-K_d/2$, given by
\begin{equation}
E_{n'_1\uparrow}(0)=E_{n'_2\downarrow}(0),n'_1=-n'_2\label{eq:degenerate-level1} 
\end{equation}
\begin{equation}
E_{n'_1\uparrow}(-\frac{K_d}{2})=E_{n'_2\downarrow}(-\frac{K_d}{2}),n'_1+n'_2=1\label{eq:degenerate-level2}
\end{equation}

Generally, the helical component $V_{n\sigma}e^{i\sigma nK_dx}$ couples $c_{n'_1,k}$ with $c_{n'_2,k}$, in which $n'_1-n'_2=n$. It is thereby quite possible that the degenerate levels listed in Eq.(\ref{eq:degenerate-level1}) and Eq.(\ref{eq:degenerate-level2}) are destroyed by one or more of the helical components. We can determine which degenerate level is broken by a certain helical component. Take the uniform field, which corresponds to $V_0$ helical component, as an example. In such case, $n'_1=n'_2=0$ and thus degeneracy of the energy level at $E_{0\uparrow(\downarrow)}(0)=0$ is destroyed, characterized with the opening of a gap at the center of FBZ.

\begin{figure}
\includegraphics[scale=0.8]{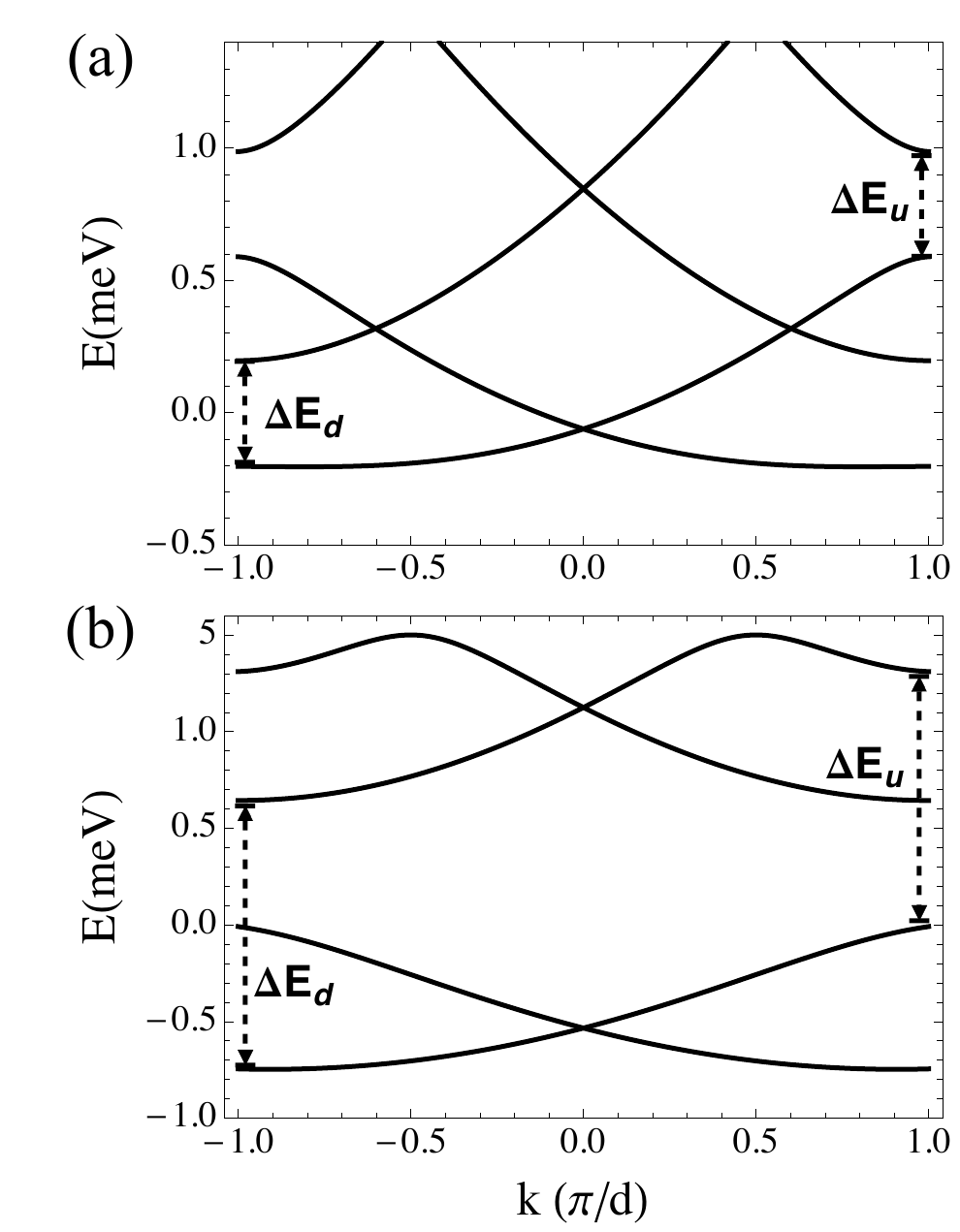}
\caption{Energy spectrum. We define a quantity with energy unit $M=\hbar^2K_d^2/(8m)$. In the plot, $M=AK_d/2=0.4$ meV. (a) Weak field case. $V_y=0.4$ meV,$V_x=0$. Two gaps $\Delta E_d$ and $\Delta E_u$ open at each FBZ edge, with equal size.  (b) Strong field case. $V_y=1.4$ meV, $V_x=0$. $\Delta E_d$ and $\Delta E_u$ overlap partially, creating a fully gapped area. \label{fig:energy-spectrum}}
\end{figure}

Returning to our case considered above, $V_+$ term in Eq.(\ref{eq:k-Zeeman}) couples $c_{0k,\uparrow}$ with $c_{1k,\downarrow}$, breaking the degeneracy at $E_{0\uparrow}(-\frac{K_d}{2})$, and giving rise to a gap denoted by $\Delta E_u$, shown in Fig. \ref{fig:energy-spectrum}. Similarly, $V_-$ term destroys degeneracy at $E_{0\downarrow}(-\frac{K_d}{2})$, being responsible for the gap denoted by $\Delta E_d$. With further analysis, we find that $V_-$ also has an influence on $\Delta E_u$, since it couples $c_{0k,\uparrow}$ with $c_{-1k,\downarrow}$. Fortunately, $V_-$ couples two non-degenerate levels and the influence on the gap size of $\Delta E_u$ is thus very limited. The same goes for $V_+$, which also has a limited effect on the gap $\Delta E_d$. Provided the Zeeman field is weak enough comparing to RSOC strength, the two gaps can be given by
\begin{equation}
\Delta E_{u(d)}=2|V_{+(-)}| \label{eq:gap}
\end{equation}

Energy spectrum in presence of the spatially varying Zeeman field is plotted in Fig.\ref{fig:energy-spectrum}, where gaps only open at the edges of FBZ. When chemical potential resides in one of these gaps, the system appears "spinless", which is the premise for the appearance of MBS. Furthermore, given that there are two separated gaps in the $E-k$ curve, MBS is expected to reappear when the chemical potential is tuned from within one gap to the other gap. One can then expect that phase transition behaviors in this system will be quite different from those in uniform fields or helical fields. In the following, we shall develop a general criterion to determine when the system reaches topological phases.

First, we perform the unitary transformation $U_-: \psi_\sigma(x)\rightarrow\psi_\sigma(x)e^{-i\sigma K_dx/2}$,\cite{Braunecker/spin-selective} and Hamiltonian in Eq.(\ref{eq:Hamiltonian}) is transformed into
\begin{eqnarray}
H&=&\sum\limits_\sigma\psi_{\sigma}^\dagger(-\frac{\hbar^2\partial_x^2}{2m}-\mu_-^{\text{eff}}-i\sigma A_-^{\text{eff}}\partial_x)\psi_\sigma \nonumber \\ 
&+&[(V_-+V_+e^{2iK_dx})\psi_{\uparrow}^\dagger \psi_\downarrow+\Delta\psi_{\uparrow}^\dagger\psi_{\downarrow}^\dagger+\text{H.c.}] \label{eq:effective-hamiltonian}
\end{eqnarray}
where
\begin{equation}
A_-^{\text{eff}}=A-\frac{\hbar^2K_d}{2m}, \mu_-^{\text{eff}}=\mu+\frac{AK_d}{2}-\frac{\hbar^2K_d^2}{8m} \label{eq:effecitive-chemical-1} 
\end{equation}
It can be seen that, after the transformation, the helical component $V_-e^{-iK_dx}$ disappears, replaced by a uniform Zeeman field term. Meanwhile, the other one $V_+e^{iK_dx}$ remains to be a helical component except the wave vector $K_d$ doubles. Without the additional oscillating term $V_+e^{2iK_dx}$,  Eq.(\ref{eq:effective-hamiltonian}) is exactly equivalent to Hamiltonian of Rashba nanowire under a uniform Zeeman field. If we could eliminate this additional term, the phase boundary can be easily obtained by comparing with that in uniform field case. However, one should be much more careful when making such bold approximations, before which we will investigate the effects of this oscillating term on the phase boundary and make sure that it won't induce qualitative changes. The detailed calculation is presented in Appendix \ref{sec:appendix_A}. Our results demonstrate that the oscillating term, in fact, only has a correction on critical chemical potential and pairing potential at the topological phase boundary, as shown in Eq.(\ref{eq:criterion_oscillating}). In the weak field case, $V_+$ is small enough compared to RSOC strength $A$, and the correction can be safely ignored. Therefore, it's reasonable to omit the oscillating term in the weak field case. This way, we can obtain the specific expression for Majorana number immediately,
\begin{equation}
\mathcal{M}=-\text{sgn}[|V_-|^2-\Delta^2-(\mu_-^{\text{eff}})^2]
\end{equation}
Topological phases, which support MBS, correspond to $\mathcal{M}=-1$,\cite{Lutchyn/multiband} yielding
\begin{equation}
|\mu_-^{\text{eff}}|<\sqrt{|V_-|^2-\Delta^2}\label{eq:criterion-1}
\end{equation} 

\begin{figure*}
\includegraphics[scale=0.55]{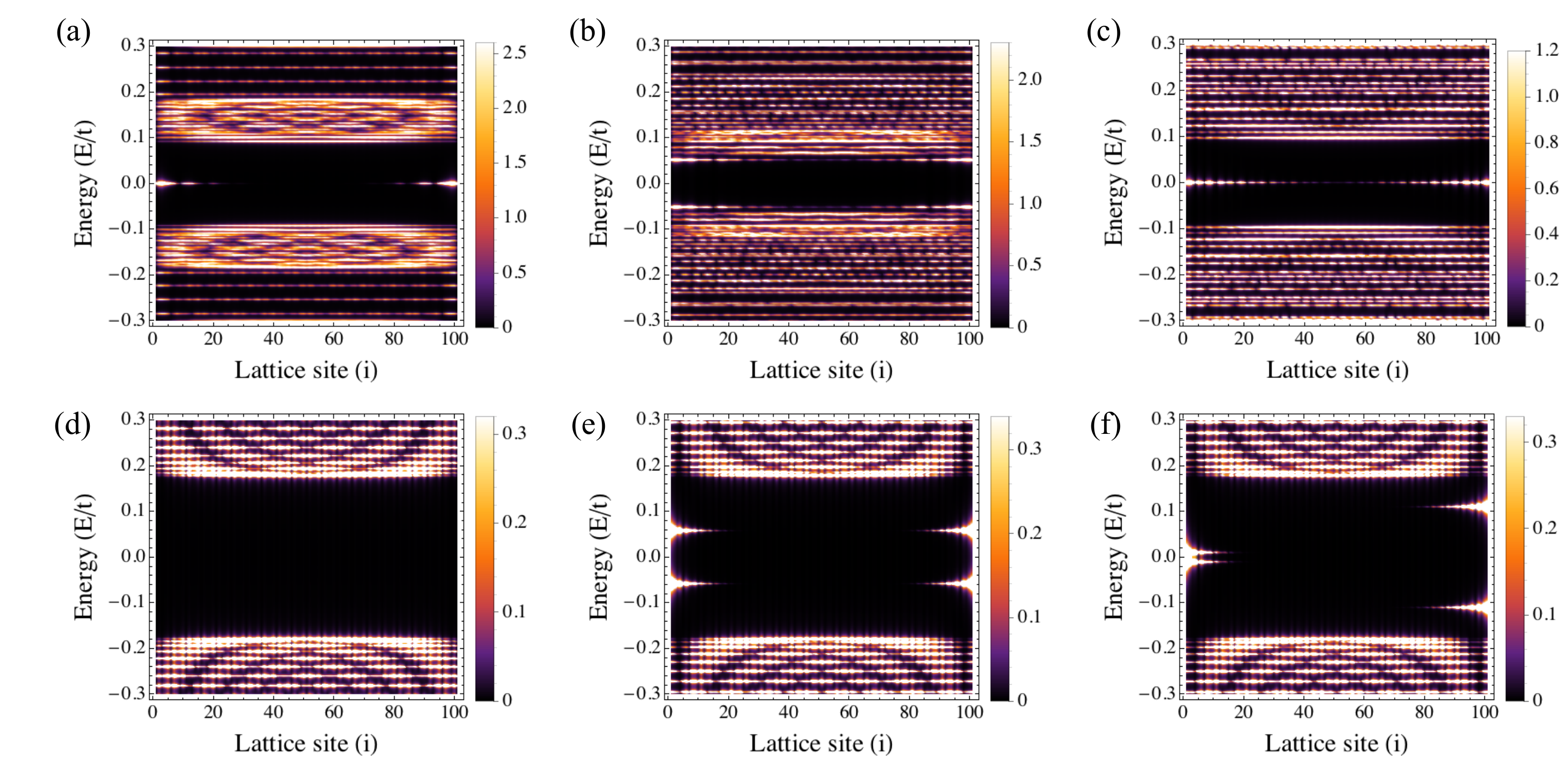}
\caption{(Color online). Local density of states. For all the plots, the period $d=5a$, lattice number $N=101$ and $A'=0.3t$. (a)-(c) $V_y=0.4t$, $V_x=0$, $\phi_y=0$,$\Delta=0.1t$. $\mu=0$, $0.3t$, $0.7t$. MBS correspond to bright lines at the center of figures). (d)-(f) $V_y=1.4t$, $V_x=0$, $\Delta=0.1t$, $\mu=0$. For (d)-(f), $\phi_y=0$, $0.5\pi$, $0.55\pi$. Energy levels of FBS are sensitive to the phases of the Zeeman field. } \label{fig:LDOS}
\end{figure*}

Similarly, we can perform another unitary transformation $U_+: \psi_\sigma(x)\rightarrow\psi_\sigma(x)e^{i\sigma K_dx/2}$. A corresponding set of effective RSOC strength and chemical potential is then given by
\begin{equation}
A_+^{\text{eff}}=A+\frac{\hbar^2K_d}{2m}, \mu_+^{\text{eff}}=\mu-\frac{AK_d}{2}-\frac{\hbar^2K_d^2}{8m} \label{eq:effecitive-chemical-2}
\end{equation}
Analogous to $U_-$, $U_+$ also generates an extra oscillating term $V_-e^{-2iK_dx}$. Provided $V_-$ is sufficiently small in comparison with RSOC strength $A$, the condition for topological phases has a similar form with Eq.(\ref{eq:criterion-1}),
\begin{equation}
|\mu_+^{\text{eff}}|<\sqrt{|V_+|^2-\Delta^2}\label{eq:criterion-2}
\end{equation}

Eq.(\ref{eq:criterion-1}) and (\ref{eq:criterion-2}) jointly determine the distribution of topological regions in the phase diagram and the system is expected to host MBS whenever either of the them is satisfied. From Eq.(\ref{eq:effecitive-chemical-1}) and Eq.(\ref{eq:effecitive-chemical-2}), one can find that effective chemical potential $\mu_\pm^{\text{eff}}$ are relevant with both chemical potential $\mu$ and RSOC strength $A$, which implies that the distribution of nontrivial phases will change with the tuning of RSOC strength. This property makes it more flexible to tune the system into topological or trivial phases as desired.

In fact, it is impossible to transform both helical components into the form of uniform fields by performing only continuous unitary transformation like $U_\pm$. As we will show in Sec. \ref{sec:numerical results}, this configuration essentially differs from the uniform field case in the sense of topology. Therefore, oscillating terms like $V_+e^{2iK_dx}$ in Eq.(\ref{eq:effective-hamiltonian}) are inevitable after performing each transformation. For fields with more than one helical components, we can perform a transformation for each of the helical components and a corresponding condition for topological phases can be reached, by ignoring the remaining oscillating terms. The general form of these criteria reads
\begin{equation}
|\mu_{n\sigma}^{\text{eff}}|<\sqrt{|V_{n\sigma}|^2-\Delta^2}\label{eq:criterion}
\end{equation}
where
\begin{equation}
\mu_{n\sigma}^{\text{eff}}=\mu-\frac{\sigma nAK_d}{2}-\frac{n^2\hbar^2K_d^2}{8m},
\end{equation}
and $|V_{n\sigma}|$ is given by Eq.(\ref{eq:Zeeman}). 

This kind of approximation works well in the weak field case, since the oscillating terms generated after each transformation only slightly shift the effective chemical and pairing potential in Eq.(\ref{eq:criterion}), as can be seen in Eq.(\ref{eq:criterion_oscillating}). The shift in the strong field case, however, will become remarkable, especially when RSOC strength is not large enough. With further investigation, we notice that in the strong field, it is quite possible that gaps induced by different helical components overlap. As illustrated in Fig. \ref{fig:energy-spectrum}(b), $\Delta E_u$ and $\Delta E_d$ overlap and creates a fully gapped area. When chemical potential resides in the overlapping area, the system has no fermion points and thus supports no MBS. Mathematically, in this case, the two inequalities (\ref{eq:criterion-1}) and (\ref{eq:criterion-2}) are satisfied simultaneously. Hence, it is reasonable to make the conclusion that the system supports no MBS when both inequalities are satisfied.

\section{numerical results}{\label{sec:numerical results}}
The analytically derived topological conditions in Sec. {\ref{sec:continuum model} basically cover the whole physics of our system in terms of topological phase transitions, especially for the case of weak Zeeman fields. In this part, we expect to make a further step and utilize numerical analysis to investigate if there could be other possible bound states inside the gap besides MBS, as well as to verify our analytical results.

A lattice model appropriate for numerical analysis can be easily translated from the continuum counterpart in Sec. {\ref{sec:continuum model}. By Fourier transforming Eq.({\ref{eq:Hamiltonian}) into $k-$ space and performing the substitution $k\rightarrow \sin k$ and $k^2\rightarrow 2(1-\cos k)$, we directly get the tight-binding Hamiltonian, which reads
\begin{eqnarray}
H=&&\sum\limits_{i\sigma\sigma'}[(-t-i\sigma A')c_{i,\sigma}^\dagger c_{i+1,\sigma}+\text{H.c.}]+(2t-\mu)c_{i\sigma}^{\dagger}c_{i\sigma}\nonumber\\
&&-[c_{i\sigma}^\dagger(\bm{V}_i\cdot \bm{\sigma})_{\sigma\sigma'} c_{i\sigma'}+\text{H.c.}]+\sigma \Delta c_{i\sigma}^{\dagger} c_{i\bar{\sigma}}^{\dagger} \label{eq:Hamiltonian_lattice}
\end{eqnarray}
where $t=\hbar^2/(2ma^2)$, is hopping parameter, and $A'=A/(2a)$, with $a$ being the lattice constant and $A$ the RSOC strength. $\bm{V}_i$, the local Zeeman field on the $i$th lattice site, has the form $\bm V_i=V_x\cos(2\pi(i-1)/n+\phi_x)\hat{\bm{e}}_x+V_y\sin(2\pi(i-1)/n+\phi_y)\hat{\bm{e}}_y$, originating from Eq.({\ref{eq:field}).

\subsection{Local density of states}{\label{subsec:LDOS}}
To investigate MBS and other possible localized states, LDOS is one of the most direct and efficient approaches.
The formula for LDOS is given by\cite{Chevallier_ABS_evolution}
\begin{equation}
n_i(\omega)=-\frac{1}{2\pi}\sum\limits_{\sigma=\uparrow,\downarrow}^N\text{Im}[G^{i\sigma,i\sigma}(\omega)] \label{eq:LDOS}
\end{equation}
where $G^-(\omega)=\omega+i\delta-H$, is Green function of the superconducting nanowire, $\sigma$ is spin index, $i$ the lattice site, and $N$ is the total number of lattice site. It should be pointed out the sum in Eq.(\ref{eq:LDOS}) is performed both for electrons and holes, which is the reason for a factor $2$ in the denominator. In the numerical calculation, we take $\delta=0.001t$ in order to avoid infinite peak values.

\begin{figure}
\includegraphics[scale=0.8]{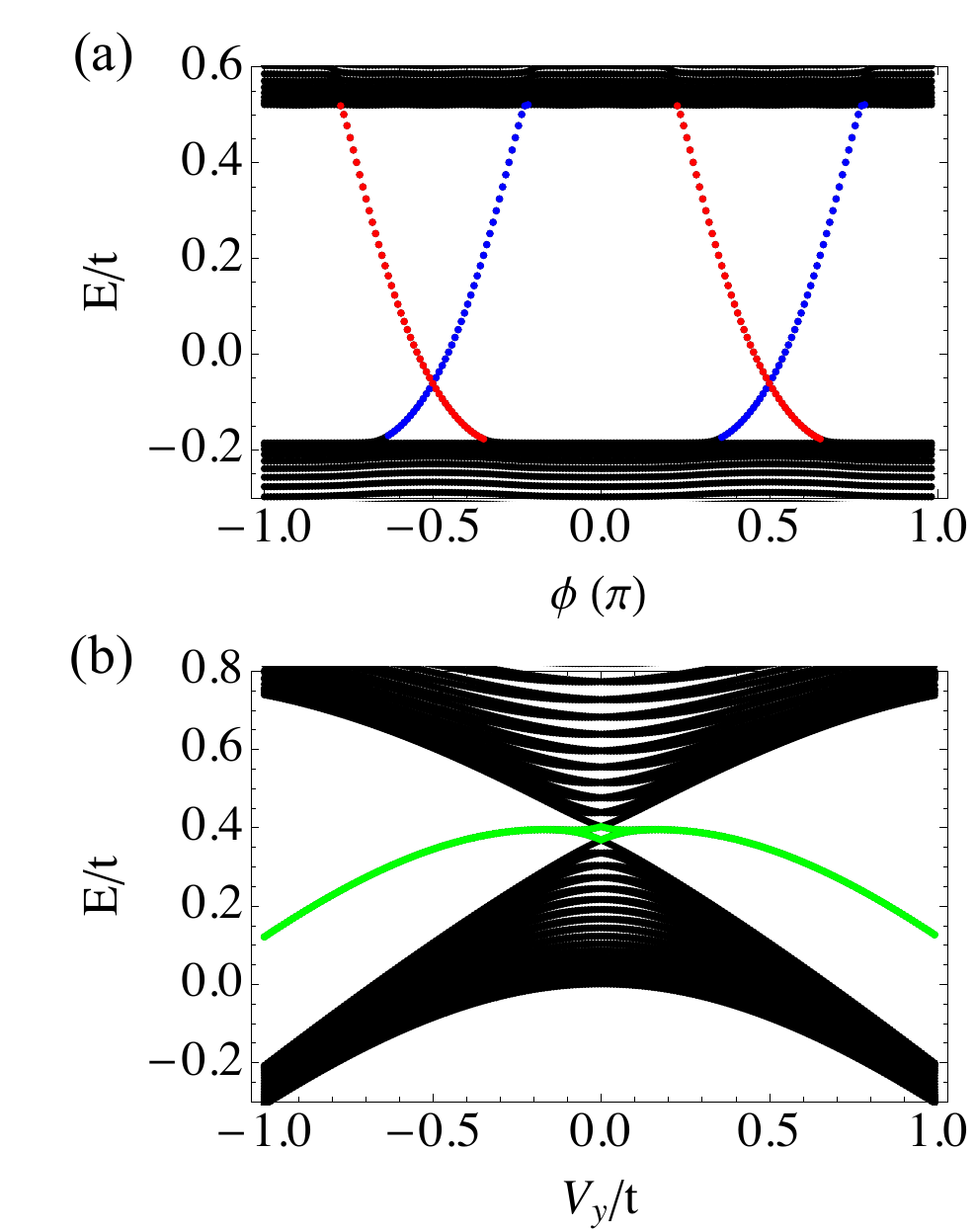}
\caption{(Color online) Bound states varying with phases and amplitude of Zeeman fields in absence of superconductivity. $N=101$, $d=5a$, $\mu=0$, $V_x=0$. (a) $V_y=1.4t$, $A'=0.3t$. The red curves represent FBS at the left end ($i=0$), whearas the blue curves represent those at the right end ($i=N$). (b) $\phi_y=\pi/2$, $A'=0t$. The green curve represents FBS, which are degenerate, since energy levels of FBS at the two ends equal to each other when $\phi_y$ takes $\pi/2$.}\label{fig:Energy_phi}
\end{figure}

As a result of splitting in the effective chemical potential, shown in Eq.(\ref{eq:effecitive-chemical-1}) and Eq.(\ref{eq:effecitive-chemical-2}), the system is expected to satisfy either of inequalities (\ref{eq:criterion-1}) and (\ref{eq:criterion-2}), or neither of them, when chemical potential varies. Therefore, one can expect the system is driven into topological trivial and nontrivial phases alternately with the increase of chemical potential. Fig. \ref{fig:LDOS} (a)-(c) illustrate the evolution of MBS with chemical potential under a weak field, where the increase of chemical potential is accompanied by the disappearance and reappearance of MBS. This exactly verifies our arguments that topological phases in a spatially varying Zeeman field are determined by a set of criteria, each of which is relating to a corresponding helical component of the field, just like Eq.(\ref{eq:criterion-1}) and (\ref{eq:criterion-2}), whose corresponding components are $V_-e^{-iK_d}x$ and $V_+e^{iK_dx}$ respectively.

In the weak field and strong RSOC regime, one needn't worry about the case where the system parameters satisfy Eq.(\ref{eq:criterion-1}) and (\ref{eq:criterion-2}) simultaneously, since the splitting extent of effective chemical potential ($|\mu_-^{\text{eff}}-\mu_+^{\text{eff}}|=AK_d$) is large compared to field strength. In the strong field case, however, we should deal with it more carefully. As illustrated in Fig. {\ref{fig:energy-spectrum}(b), two gaps $\Delta E_d$ and $\Delta E_u$ overlap partially. When chemical potential resides in the overlapping region, $i.e.$, the fully gapped area, it's quite possible the system parameters fulfill both criteria shown in Eq.(\ref{eq:criterion-1}) and (\ref{eq:criterion-2}). In this circumstance, the system supports no MBS, as we can see in Fig. {\ref{fig:LDOS}(d). Instead, other bound states may appear, the energy levels of which vary with phases of the Zeeman field, as Fig. {\ref{fig:LDOS}(e)-(f) shows. Interestingly, these localized states inside the gap survive in the absence of superconductivity, being plotted in Fig. \ref{fig:Energy_phi}. A distinctive property of these states is that, the energy levels behave sensitively to phases of the field, contrasting to MBS, which are immune to the variance of phases. 

This kind of states arises from the overlap of two gaps with nontrivial topology. As the two gaps are induced by different helical components, which have distinguishing topological configurations, it's expected that these two gaps possess different topology. Analogous to SSH model,\cite{SSH,Fractional_Su,Fractional_Goldstone} where fermionic bound states carrying fractional charge are induced at the boundaries separating two topologically different configurations, these localized bound states are also fractionally charged bound states (FBS) of Jackiw-Rebbi type.\cite{Jackiw-Rebbi} Similar states have been studied in Ref.[\onlinecite{fractionalfermions}], in which FBS emerge in the overlapping zone of a helical field induced gap and a uniform field induced one, and in our context, the uniform field exactly corresponds to $V_{n=0}$ helical component in Eq.(\ref{eq:Zeeman}). Instead, FBS shown in Fig. {\ref{fig:LDOS}(e)-(f), originate from the interplay between $V_+$ and $V_-$ components.

Despite that different helical components can be transformed into each other by performing transformations like $U_\pm$, they still represent different topological configurations, if one notice that such transformations inevitably revise chemical potential and RSOC strength. Therefore, for a system with fixed chemical potential and RSOC, different helical components applied have distinguishing topological origin and thus the overlapping gapped area induced by relevant helical components can possibly support FBS like that formed at the boundaries of two different configurations in SSH model. In this sense, one cannot transform a field with two helical components into a uniform one, just like the knot or anti-knot in SSH model cannot be disentangled by continuous transformation.

Briefly speaking, both MBS and FBS can possibly appear under spatially varying Zeeman fields. MBS emerge in the gap with nontrivial topology, while FBS in the overlapping zone of two gaps possessing different topology. We can make the conclusion that, in weak field and strong RSOC case, the system favors MBS, since gaps relating to different helical components can hardly overlap, whereas FBS are more favorable in the opposite case. Indeed, systems with weak RSOC strength or even without RSOC favor FBS, in which circumstance, $\mu_-^{\text{eff}}-\mu_+^{\text{eff}}=0$ and thereby FBS occur even under a weak enough field, as shown in Fig. \ref{fig:Energy_phi}(b).

\subsection{Phase diagram}{\label{subsec:phase diagram}}
\begin{figure}
\includegraphics[scale=0.55]{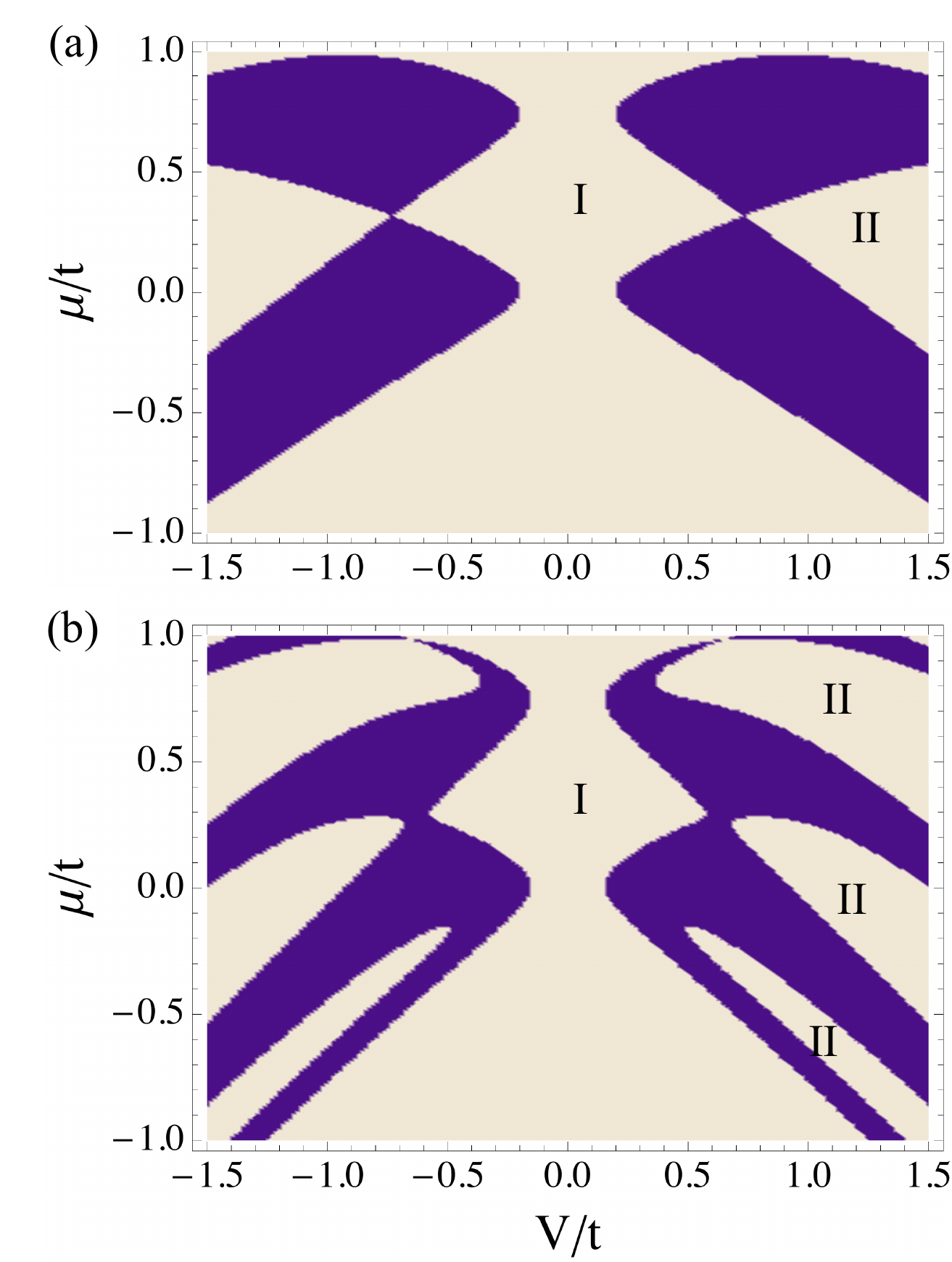}
\caption{(Color online). Phase diagram. $\Delta=0.1t$, $d=5a$. Blue area represent topological phases which host MBS, the light yellow regions denoted by phase-II support FBS and the remaining part (phase-I) supports neither MBS nor FBS. (a) In a field with sinusoidal form. $\bm V_i=V\sin[2\pi(i-1)/5]\hat{e}_y$. (b) In a field of square wave form, $\bm V_i=V\hat{e}_y$ for $i=5m, 5m+1, 5m+2$, and $\bm V_i=-V\hat{e}_y$ for $i=5m+3, 5m+4$. In plot (b), the displaying three phase-II regions arise from the interplay between $n=0$ and $1$, $n=1$ and $-1$, $n=-1$ and $2$ helical components, respectively.}\label{fig:phase-diagram}
\end{figure}

To determine phase boundaries, we compute Majorana number using the formulation proposed by Kitaev.\cite{Kitaev} We first write the tight-binding Hamiltonian (\ref{eq:Hamiltonian_lattice}) in Majorana basis, which consists of all sites in a lattice period and calculate the Pfaffian of the resulting anti-symmetric matrix. Periodic boundary condition is assumed in our calculation. Note the period of the 1-D lattice under spatially varying Zeeman fields is $d=na$, rather than the original lattice period $a$. Therefore, the Pffafin calculation should be performed in the super cell consisting of $2n$ fermionic sites, with spin components involved.

For a field with only two helical components, the phase diagram are supposed to be divided into three types of regions, representing differing phases, just as Fig. \ref{fig:phase-diagram}(a) shows. Regions that satisfy either of the criteria in Eq.(\ref{eq:criterion-1}) and (\ref{eq:criterion-2}) (blue colored regions in Fig. \ref{fig:phase-diagram}(a)) support MBS, that satisfy both of them (phase-II in light yellow) support FBS, and otherwise neither MBS nor FBS (phase-I). More regions of phase-II are expected if the system are subject to a field with more than two helical components, since regions in phase diagram fulfilling any two conditions of Eq.(\ref{eq:criterion}) are expected to be in phase-II, as Fig. \ref{fig:phase-diagram}(b) demonstrates, where a Zeeman field with square wave form is applied.

The set of topological criteria in Eq.(\ref{eq:criterion}) can be used to determine phase distribution of our system, except that the corrections on chemical and pairing potential need to be considered in presence of strong fields. Regions in the phase diagram which satisfy any two of the criteria support FBS, those satisfy only one of them support MBS, and those satisfy none of them support no bound states. It should be noted that FBS survive in absence of superconductivity, in which case, however, MBS disappear. As the effective chemical potential in formula \ref{eq:criterion} depends on RSOC strength, the phase distribution can be adjusted more flexibly. We can thereby choose a optimal phase distribution where the system can enter into the desired phases easily and be immune to fluctuations of chemical potential in a broad range.
\subsection{Spin texture}{\label{subsec:spin texture}}
\begin{figure*}
\includegraphics[scale=0.4]{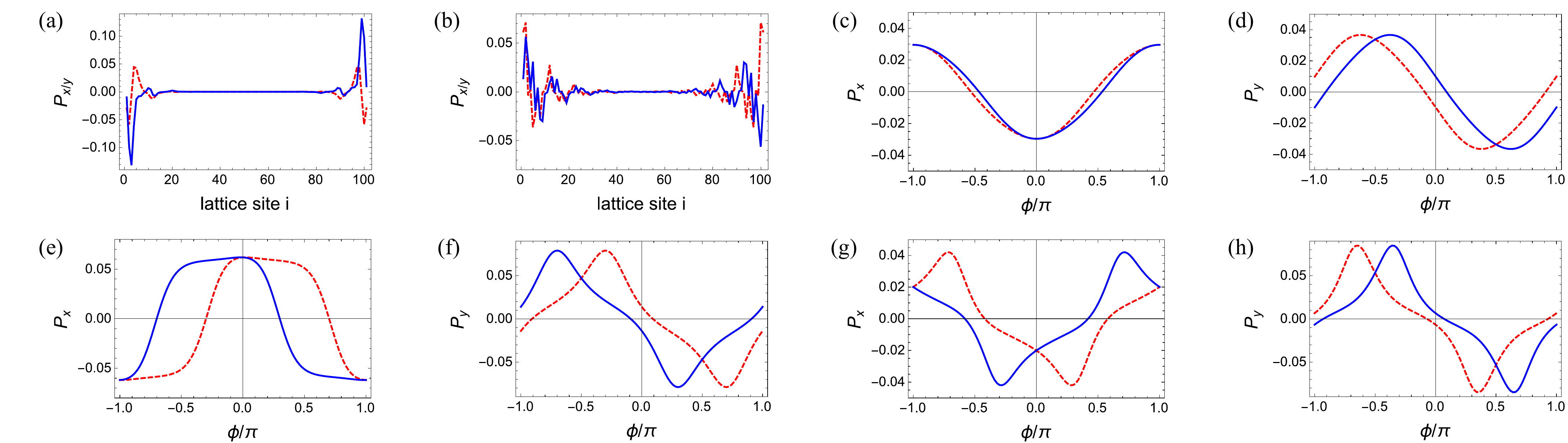}
\caption{(Color online). Spin polarization of zero energy bound states. The period $d=5a$, lattice number $N=101$, $\Delta=0.1t$, $A'=0.3t$, $V_x=0$. (a) and (b) Spin polarization decays with different charateristic length. Red (dashed) line represent $P_x$, blue (solid) line $P_y$. $V_y=0.4t$. $\mu=0$ for (a) and $\mu=0.75t$ for (b). (c)-(h) Spin polarization at both ends vary with phases of the field. Red (dashed) line represents spin polarization at the first lattice point and blue (solid) line the last lattice point. $V_y=0.4t$, $\mu=0$ for (c) and (d). $V_y=0.4t$, $\mu=0.75t$ for (e) and (f). $V_y=t$, $\mu=0$ for (g) and (h).}\label{fig:spin-texture}
\end{figure*}
Finally, we study the spin texture of zero energy bound states in topological phases. It has been shown that, in presence of uniform fields, spin components perpendicular to RSOC vector are polarized, whereas the polarization along RSOC vector is zero.\cite{Sticlet/spin-polarization} We shall investigate its behaviors under spatially varying Zeeman fields. $\sigma_n$ component of local spin polarization at the $j$th site is given by \cite{Sticlet/spin-polarization}
\begin{equation}
s_{j,n}=\langle \Psi_j\vert\sigma_n\frac{\tau_0+\tau_z}{2}\vert\Psi_j\rangle
\end{equation}
where $\vert\Psi_j\rangle=\{u_{j\uparrow},u_{j\downarrow},v_{j\downarrow},-v_{j\uparrow}\}$, is the $j$th component of the eigenvectors and $n=x,y,z$, representing three spin components.

As Fig. \ref{fig:spin-texture}(a) and (b) illustrate, $\sigma_x$ component is also polarized even if the Zeeman field is arranged along $y$ direction. This is because each topological region, as we have demonstrated, is relating to a helical component of the Zeeman field. Take the helical component $V_-e^{-iK_dx}$ for example. By rewriting it as $V_-[\cos(k_dx)+i\sin(K_dx)]$, we can easily find that this Zeeman term is actually equivalent to a Zeeman field with both $x$ and $y$ components. So, under the spatially varying field, both spin components perpendicular to the RSOC vector are supposed to be polarized. It happens also in presence of the uniform field, as investigated in Ref.[\onlinecite{Sticlet/spin-polarization}], where both spin components perpendicular to RSOC vector polarizes. This can be understood if one notice that a uniform field can be transformed into a helical field in the plane perpendicular to RSOC and thereby they share the same property. That's why we can include the uniform field as one helical component. Fig. \ref{fig:spin-texture}(a) and (b) belong to different topological regions, $i.e.$, regions determined by differing helical components.  

In Appendix \ref{sec:appendix_B}, we derive the spin polarization of zero mode states in a field with only one helical component. As can be seen in Eq.(\ref{eq:spin-polarization-1}) and Eq.(\ref{eq:spin-polarization-2}), spin polarization at both ends are the cosine or sine functions of phases of the Zeeman field, exactly as Fig. \ref{fig:spin-texture}(c)-(d) display. Despite MBS are immune to variance of phases, we find spin polarization profile at zero energy in topological phases, which relates to the wave function of MBS, oscillates remarkably with phases. 

In the strong field, $P-\phi$ curves shown in Fig. \ref{fig:spin-texture}(g)-(h) are seriously distorted compared with cosine and sine curves, arising from strong interplay between different helical components. Interestingly, at some special points, the relation of polarization for the two ends keeps unchanged. As can be seen in Fig. \ref{fig:spin-texture}(g)-(h), the following relations hold for any topological regions.
\begin{eqnarray}
P_x(0)&=&P_x(L), P_y(0)=-P_y(L), \phi=0,\pi\nonumber\\
P_x(0)&=&-P_x(L), P_y(0)=P_y(L), \phi=\pm\pi/2,
\end{eqnarray}
where $P_{x/y}(0/L)$ represents polarization of spin $x$ and $y$ component at the two ends of nanowire, $x=0,L$. This relations can be well explained by Eq.(\ref{eq:spin-polarization-1}) and Eq.(\ref{eq:spin-polarization-2}), in which spin polarization for different topological regions actually share the same relations. As a result, interplay between different helical components have no influences on them.

\section{summary}\label{sec:summary}
In summary, we show that when a 1-D Rashba nanowire is subject to spatially varying Zeeman fields, two kind of localized states, MBS and FBS, can emerge. Unlike the case in uniform fields or helical fields with only one helical component, there exist multiple topological regions in our system, each of which is relating to one helical component of the field and can be described by a separate topological criterion. Systems that satisfy one of the criteria locates in topological phases and support MBS, those satisfying any two of the criteria support FBS, energy levels of which behave sensitively with phases of the Zeeman field, and those satisfying none of the criteria support no bound states. Based on these criteria, we find that systems with strong RSOC and weak Zeeman fields favor MBS, whereas FBS are favorable in the opposite case. Moreover, we demonstrate that FBS results from the interplay of any two helical components with topologically different configurations, just as that formed at the knot in SSH model. In the end, we investigate local spin polarization of zero mode states and show that only spin perpendicular to RSOC vector are polarized. In addition, spin polarization changes remarkably with variance of phases of the Zeeman field, indicating that wave functions of MBS behave sensitively to phases.

\appendix
\section{Effects of oscillating terms on the topological phase boundary} \label{sec:appendix_A}
In this appendix, we shall deal with the additional oscillating term in Eq.(\ref{eq:effective-hamiltonian}), and focus on its effects on the topological phase boundary.

First, we write Eq.(\ref{eq:effective-hamiltonian}) in the Nambu basis $\{ \psi_\uparrow(x),\psi_\downarrow(x),\psi^\dagger_\downarrow(x),-\psi^\dagger_\uparrow(x) \}$, and the BdG Hamiltonian matrix reads
\[
H_{BdG}=\left(
\begin{array}{ccc}
  H_p & \Delta\\
  \Delta & H_{p-h} 
\end{array}
\right)
\]

\[
H_p=\left(
\begin{array}{ccc}
  -\frac{\hbar^2\partial_x^2}{2m}-\mu_-^{\text{eff}}-iA_-^{\text{eff}}\partial_x & V_-+V_+e^{2iK_dx}\\
 V_-^*+V_+^*e^{-2iK_dx} & -\frac{\hbar^2\partial_x^2}{2m}-\mu_-^{\text{eff}}+iA_-^{\text{eff}}\partial_x
\end{array}
\right)
\]

\[
H_{p-h}=\left(
\begin{array}{ccc}
  \frac{\hbar^2\partial_x^2}{2m}+\mu_-^{\text{eff}}+iA_-^{\text{eff}}\partial_x & V_-+V_+e^{2iK_dx}\\
 V_-^*+V_+^*e^{-2iK_dx} & \frac{\hbar^2\partial_x^2}{2m}+\mu_-^{\text{eff}}-iA_-^{\text{eff}}\partial_x
\end{array}
\right)
\]

Without the oscillating term $V_+e^{2iK_dx}$, the wave function of this Hamiltonian has a general form, $\Psi(x)=\{u,v\}^Te^{ikx}$, and $u,v$ are two-component spinors, which are independent of position $x$. To decide whether the system is located in topological nontrivial phases or not, we only need to solve the BdG equation $H\Psi=0$ and to see if it has a solution where $k$ has an imaginary part. The phase boundary is determined by substituting $\Psi(x)$ into the equation, and solving $\det (H_{k=0})=0$. In this case, 
\[
H_{k=0}=\left(
\begin{array}{cccc}
-\mu_-^{\text{eff}} & V_- & \Delta & 0 \\
  V_-^* & -\mu_-^{\text{eff}} & 0 & \Delta \\
  \Delta & 0 & \mu_-^{\text{eff}} & V_- \\
  0 & \Delta & V_-^* & \mu_-^{\text{eff}}\\
\end{array}
\right)
\]
Solving $\text{Det}(H_k)=0$, we obtain the formula of the phase boundary, given by $|\mu_-^{\text{eff}}|=\sqrt{|V_-|^2-\Delta^2}$.

With this oscillating term, the wave function should be revised as 
\begin{equation}
\Psi(x)=\sum\limits_n\{u_ne^{2inK_dx},v_ne^{-2inK_dx}\}^Te^{ikx}
\end{equation}
By substituting it into BdG equation and adding the terms with the same $e^{2inK_dx}$ together, we will get numerous equations, which read
\begin{equation}
H_{k=2nK_d}\left(
\begin{array}{ccc}
  u_n   \\
  v_{-n}    \\ 
\end{array}
\right)+
\hat{V}_+\left(
\begin{array}{ccc}
  u_{n-1}   \\
  v_{-n+1}    \\ 
\end{array}
\right)+
\hat{V}_+^\dagger \left(
\begin{array}{ccc}
  u_{n+1}   \\
  v_{-n-1}    \\ 
\end{array}
\right)=0 \label{eq:n_BdG_equation}
\end{equation}
where $n$ takes any integer. 
\[
H_{k}=\left(
\begin{array}{cccc}
 M-\tilde{A} & V_-  & \Delta & 0 \\
  V_-^* & M+\tilde{A}  & 0 & \Delta \\
  \Delta & 0   &   -M+\tilde{A} & V_-\\
  0 & \Delta & V_-^* & -M-\tilde{A}  
\end{array}
\right)
\]

$M=M_k=\frac{\hbar^2k^2}{2m}-\mu_-^{\text{eff}}$, and $\tilde{A}=\tilde{A}_k=A_-^{\text{eff}}k$

\[
\hat{V}_+=\left(
\begin{array}{cccc}
 0 & V_+ & 0 & 0  \\
 0 & 0  & 0 & 0 \\
 0 & 0  & 0 & V_+\\
 0 & 0 & 0 & 0
\end{array}
\right)
\]
It's quite a complicated work to solve Eq.(\ref{eq:n_BdG_equation}) when $n$ takes an infinite set of integers. Fortunately, $\{u_n,v_{-n}\}^T$ actually decrease with the increase of $n$, since the diagonal elements of the coefficient matrix $H_{2nK_d}$ increase with $n$. So, only components of small $n$ play the leading role. We can then only consider the wave function with $n=0,\pm 1$ components and the resulting three equations read,
 
\[
H_{k=0}\left(
\begin{array}{ccc}
  u_0   \\
  v_0    \\ 
\end{array}
\right)+
\hat{V}_+\left(
\begin{array}{ccc}
  u_{-1}   \\
  v_{1}    \\ 
\end{array}
\right)+
\hat{V}_+^\dagger \left(
\begin{array}{ccc}
  u_{1}   \\
  v_{-1}    \\ 
\end{array}
\right)=0
\]
\[
H_{k=2K_d}\left(
\begin{array}{ccc}
  u_1   \\
  v_{-1}    \\ 
\end{array}
\right)+
\hat{V}_+\left(
\begin{array}{ccc}
  u_{0}   \\
  v_{0}    \\ 
\end{array}
\right)=0
\]
\[
H_{k=-2K_d}\left(
\begin{array}{ccc}
  u_{-1}   \\
  v_1    \\ 
\end{array}
\right)+
\hat{V}_+^\dagger \left(
\begin{array}{ccc}
  u_{0}   \\
  v_{0}    \\ 
\end{array}
\right)=0
\]
Representing $\{ u_{\pm 1},v_{\mp 1} \}$ with $\{ u_0,v_0 \}$, we get 
\[
(H_{k=0}-H')\left(
\begin{array}{cc}
  u_0 \\
  v_0   
\end{array}
\right)=0
\]
where 
\begin{equation}
H'=\hat{V}_+^\dagger H^{-1}_{k=2K_d}\hat{V}_+ -\hat{V}_+ H^{-1}_{k=-2K_d} \hat{V}_+^\dagger \nonumber
\end{equation}
$H'$ can be written explicitly as follows
\[
H'=\left(
\begin{array}{cccc}
 \delta \mu & 0  & \delta \Delta & 0  \\
  0 &   \delta \mu & 0 & \delta \Delta  \\
  \delta \Delta & 0  &  -\delta \mu & 0 \\
  0 & 0 & \delta \Delta &  -\delta \mu   
\end{array}
\right)
\]
where
\begin{eqnarray}
\delta \mu &=& \frac{|V_+|^2}{N}[\tilde{A}_{2K_d}(T+|V_-|^2)+M_{2K_d}(T-|V_-|^2)] \nonumber \\
\delta \Delta &=&\frac{|V_+|^2\Delta}{N}(T-|V_-|^2) \\
T &=& (\tilde{A}_{2K_d}-M_{2K_d})^2+\Delta^2 \nonumber \\
N &=& [2\tilde{A}_{2K_d}(\tilde{A}_{2K_d}-M_{2K_d})+|V_-|^2-T]^2+4\tilde{A}^2_{2K_d}\Delta^2 \nonumber
\end{eqnarray}
So the oscillating term in fact shift the chemical potential as well as pairing potential. By solving $\det(H_{k=0}-H')=0$, we get the revised topological phase boundary, given by
\begin{equation}
|\mu_-^{\text{eff}}+\delta \mu|=\sqrt{|V_-|^2-(\Delta-\delta \Delta)^2}\label{eq:criterion_oscillating}
\end{equation}
Actually, in the weak field case, we can completely ignore this correction term, which is important only when the Zeeman field is comparable to RSOC strength.

\section{Spin polarization of Majorana bound states in a helical Zeeman field}\label{sec:appendix_B}
In this appendix, we only consider the field with only one helical component. As we have demonstrated in the main text, each topological region is relating with a helical component. So, we can analyze the spin polarization of MBS in any one of the topological regions by considering mainly the corresponding helical component. Take the field $\vec{V}(x)=V\cos(K_dx+\phi)\bm{\hat{e}}_x+V\sin(K_dx+\phi)\bm{\hat{e}}_y$ for example. The Zeeman Hamiltonian is then given by 
\begin{equation}
H_Z(x)=Ve^{-iK_dx-i\phi}\psi^\dagger_\uparrow(x)\psi_\downarrow(x)+\text{H.c.}
\end{equation}
The eigenvectors of MBS have the general form
\begin{equation}
\Psi^T=\{e^{-i\alpha(x)}u_\uparrow,e^{i\alpha(x)}u_\downarrow,e^{-i\alpha(x)}v_\downarrow,-e^{i\alpha(x)}v_\uparrow\}\label{eq:wavefunction}
\end{equation}
where $\alpha(x)=(K_dx+\phi)/2$. Substituting the eigenvector into the BdG equation $H\Psi=0$, we obtain 
\begin{equation}
H_-^{\text{eff}}\bm u=0, \bm u^T=\{u_\uparrow,u_\downarrow,v_\downarrow,-v_\uparrow\} \nonumber 
\end{equation}
\[
H_-^{\text{eff}}=\left(
\begin{array}{ccc}
  H_p & \Delta\\
  \Delta & H_{p-h} 
\end{array}
\right)
\]

\[
H_p=\left(
\begin{array}{ccc}
  -\mu_-^{\text{eff}}-iA_-^{\text{eff}}\partial_x & V\\
 V & -\mu_-^{\text{eff}}+iA_-^{\text{eff}}\partial_x
\end{array}
\right)
\]

\[
H_{p-h}=\left(
\begin{array}{ccc}
  \mu_-^{\text{eff}}+iA_-^{\text{eff}}\partial_x & V\\
 V & \mu_-^{\text{eff}}-iA_-^{\text{eff}}\partial_x
\end{array}
\right)
\]
where we keep only the linear part. The effective Hamiltonian is written in the basis $\{ \psi_\uparrow(x),\psi_\downarrow(x),\psi^\dagger_\downarrow(x),-\psi^\dagger_\uparrow(x) \}$. We assume the region is topological non-trivial in $0<x<L$, and trivial in others. We use the wave functions provided in Ref.[\onlinecite{Sticlet/spin-polarization}], with a little revision in eigenvectors, given by 
\begin{eqnarray}
\bm u_1^T&=&\frac{1}{2}\{ie^{-i\theta},-i,e^{-i\theta},1\} \nonumber \\
\bm u_2^T&=&\frac{1}{2}\{-ie^{-i\theta},i,e^{-i\theta},1\} \nonumber \\
\bm u_3^T&=&\frac{1}{2}\{-ie^{i\theta},i,e^{i\theta},1\} \nonumber \\
\bm u_4^T&=&\frac{1}{2}\{ie^{i\theta},-i,e^{i\theta},1\}
\end{eqnarray}
where $e^{i\theta}=(\mu_-^{\text{eff}}-i\sqrt{V^2-(\mu_-^{\text{eff}})^2})/V$. All of the vectors above could be transformed into the form $\{u,v,v^*,-u^*\}$ by multiplying by a global phase, indicating they are indeed the Majorana vectors. Utilizing the formula $s_i(x)=\langle\psi|\sigma_i\frac{\tau_0+\tau_z}{2}|\psi\rangle$, we obtain the spin polarization for Majorana bound states, which reads
\begin{eqnarray}
\bm s(0)&=&\frac{C}{2}(-\cos(\phi+\theta),-\sin(\phi+\theta),0)\\ \label{eq:spin-polarization-1}
\bm s(L)&=&\frac{C}{2}(-\cos(\phi+K_dL-\theta),-\sin(\phi+K_dL-\theta),0)\nonumber
\end{eqnarray} 
where $C$ is the normalization coefficient given by $C\int dx \exp(-2\kappa x)=1$, $\kappa=(\Delta-\sqrt{V^2-(\mu_-^{\text{eff}})^2})/A_-^{\text{eff}}$.
These results are derived under the assumption that $A_-^{\text{eff}}<0$. Similarly, we can write spin polarization for $A_-^{\text{eff}}>0$, as follows
\begin{eqnarray}
\bm s(0)&=&\frac{C}{2}(-\cos(\phi+\theta),\sin(\phi+\theta),0)\\ \label{eq:spin-polarization-2}
\bm s(L)&=&\frac{C}{2}(-\cos(\phi+K_dL-\theta),\sin(\phi+K_dL-\theta),0)\nonumber 
\end{eqnarray} 
Notice the sign of y-component polarization is reversed comparing with $A_-^{\text{eff}}<0$ case.

\end{document}